\documentstyle[12pt]{article}
 \def\be{\begin{equation}}
 \def\ee{\end{equation}}
 \def\bea{\begin{eqnarray}}
 \def\eea{\end{eqnarray}}
 \def\nn{\nonumber}
 \def\vx{{\vec x}}
 \def\vk{{\vec k}}
 \def\vy{{\vec y}}
 \def\vz{{\vec z}}
 \def\p{\psi}
 \def\pb{\bar{\psi}}
 \def\d{\not{\!\!D}}
 \def\tp{\tilde{\psi}}

\begin{document}
 \title{Massive Spinors and dS/CFT Correspondence}
 \author{Farhang Loran\thanks{e-mail:
loran@cc.iut.ac.ir}\\ \\
  {\it Department of  Physics, Isfahan University of Technology (IUT)}\\
{\it Isfahan,  Iran,} \\
  {\it Institute for Studies in Theoretical Physics and Mathematics (IPM)}\\
{\it P. O. Box: 19395-5531, Tehran, Iran.}}
\date{}
\maketitle

 \begin{abstract} Using the map between free massless spinors on d+1
 dimensional Minkowski spacetime and free massive spinors on
 $dS_{d+1}$,  we obtain the boundary term that should be added to the standard Dirac action
 for spinors in the dS/CFT correspondence. It is shown that this
 map can be extended only to theories with vertex $({\bar\p}\p)^2$ but arbitrary
 $d\ge1$. In the case of scalar field
 theories such an extension can be made only for $d=2,3,5$ with
 vertices $\phi^6$, $\phi^4$ and $\phi^3$ respectively.
\end{abstract}

\newpage
 \section{Introduction}
 It is known that the correct action for spinors in
 $AdS_{d+1}/CFT_d$  correspondence is the sum of the standard Dirac action,
 \be
 S=\int_\Sigma d^{d+1}x \pb(x)\left(\d-m\right)\p(x),
 \label{a1}
 \ee
 and some boundary term \cite{Hen,Asl,ref}. Specially in \cite{Hen}, Henneaux  has shown that
 the boundary term can be determined by the stationary conditions on the solutions of
 Dirac equation with a definite asymptotic behavior. In those
 papers, the Euclidean AdS space is considered as the domain $t>0$ with
 metric,
 \be
 ds^2=\frac{1}{t^2}\eta^{\mu\nu}dx^\mu dx^\nu,
 \label{metric}
 \ee
 where $x^0=t$ and $\eta^{\mu\nu}=(+,+,\cdots,+)$. It is also shown that the boundary term
 gives the action for spinors on the boundary with expected
 conformal weight in AdS/CFT correspondence. Since, $dS_{d+1}$ space can be also
 described as the domain $t>0$ with metric (\ref{metric}) in which
 $\eta^{\mu\nu}=(+,-,\cdots,-)$, it is reasonable to consider
 similar boundary terms in dS/CFT correspondence.
 \par
 Recently we showed that massless scalar fields on $d+1$ dimensional Minkowski
 spacetime are dual to scalar fields on $dS_{d+1}$ with mass
 $m^2=\frac{d^2-1}{4}$ \cite{map}. In fact, if $\phi(\vx,t)$ is the solution of Klein-Gordon
 equation in $d+1$ dimensional Minkowski spacetime $M_{d+1}$,
 \be
 \eta^{\mu\nu}\partial_\mu\partial_\nu\phi(t,\vx)=0,
 \label{MKG}
 \ee
 then $\Phi(\vx,t)=t^{\frac{d-1}{2}}\phi(\vx,t)$ is a massive scalar on $dS_{d+1}$,
 satisfying the Klein-Gordon equation,
 \be
 \left(t^2\partial_t^2+(1-d)t\partial_t-t^2\nabla^2+\frac{d^2-1}{4}\right)\Phi(\vx,t)=0
 \label{dSKG}
 \ee
 Using this map we obtained the action of the dual CFT on
 the boundary of dS space given in \cite{Strom} by inserting the solution of the Klein-Gordon
 equation in $M_{d+1}$ in terms of the initial data given on
 the hypersurface $t=0$ into the action of massless scalars on
 $M_{d+1}$ and identifying the initial data with the CFT
 fields. The same result was obtained by mapping massless scalars from
 Euclidean space to Euclidean AdS.
 \par
 In $M_{d+1}$, solutions of Dirac equation are solutions of
 the Klein-Gordon equation. Therefore, one can obtain
 some information about spinors and the dS/CFT correspondence by
 mapping massless spinors from $d+1$ dimensional Minkowski (Euclidean) spacetime to
 de Sitter (Euclidean AdS) space.
 \par
 As is mentioned in \cite{map}, one can use the relation between massless fields in
 Minkowski (Euclidean) spacetime and massive fields in dS  (Euclidean AdS) space
 to prove the holographic principle for the domain $t\ge0$ of Minkowski (Euclidean) spacetime
 in the case of massless fields. In the case of Minkowski spacetime a covariant description
 of the holographic principle can be given by considering a covariant boundary instead of the
 hypersurface $t=0$ which here is the space-like boundary of the domain $t\ge0$
 \cite{Bus}. Similar ideas are considered in ref.\cite{Sol} where the relation
 between massive and massless scalar fields in Minkowski spacetime and on (anti-)de Sitter
 space, one dimension lower, is studied.
 \par
 The organization of paper is as follows. In section 2, we give
 the solution of Dirac equation for  massless spinors in the domain $t\ge0$ of Minkowski
 spacetime in terms of the initial data given on the boundary $t=0$.
 Specially we show that by
 adding a suitable  boundary term to the standard Dirac action for spinors,
 not only the Hermitian condition of the action can be satisfied but one can also obtain the
 Dirac equation without imposing any condition on fields living on the boundary.
 Finally we will consider this unique boundary term as the action of
 CFT on the boundary. In section 3, we show that the most general
 solution of Dirac equation on dS (Euclidean AdS) space (spinors with arbitrary mass) can be
 given in terms of massless spinors in Minkowski spacetime with
 the same dimensionality.  Finally we show that
 the map between fields in Minkowski spacetime and dS space can be
 extended to field theories with vertex $({\bar\p}\p)^2$ but
 arbitrary $d\ge 1$ in the case of spinors. For scalar field
 theories this extension is applicable only in $d=2,3,5$ and for
 theories with vertices $\phi^6$, $\phi^4$ and $\phi^3$
 respectively. We close the paper with a brief summary of results.
 There is also an appendix in which we have given some details
 about the general solution of Dirac equation for massless spinors
 in $M_{d+1}$. These details are considerable since they show the
 agreement of our results with those of dS/CFT in the case of
 scalar fields reported in \cite{map,Strom}. In addition they make
 the connection of our results with previous attempts on dS/CFT in
 the case of spinors \cite{Hen,Asl,ref}.

 %------------------------------------------------------------------------
 \section{Massless Fermions on $M_{d+1}$}
 The Dirac equation for massless fermions, $i\gamma^\mu\partial_\mu\psi(t,\vx)=0$ has
 solutions,
 \be
 \p(\vx,t)=\int d^d\vk \tp^\pm(\vk) e^{i\vk.\vx}e^{\pm i\omega t},
 \label{b1}
 \ee
 where,
 \be
 \left(\pm \omega\gamma^0+\gamma^i\vk_i\right)\tp^\pm(\vk)=0,
 \hspace{1cm}\omega=\left|\vk\right|.
 \ee
 $\gamma^0$ and $\gamma^i$'s are Dirac matrices
 satisfying the anti-commutation relation
 $\{\gamma^\mu,\gamma^\nu\}=2\eta^{\mu\nu}$, in which
 $\eta^{\mu\nu}=(+,-,\cdots,-)$. One can show that
 $\tp^+=\pm\gamma^0\tp^-$, therefore a general solution, $\p(\vx,t)$ of the Dirac equation
 can be decomposed as $\p(\vx,t)=\p^{(1)}(\vx,t)+\p^{(2)}(\vx,t)$, where
 \bea
 \p^{(1)}(\vx,t)&=&\int d^d\vk \left(e^{-i\omega t}+\gamma^0 e^{i\omega t}\right)
 \tp^{(1)}(\vk)
 e^{i\vk.\vx},\nn\\
 \p^{(2)}(\vx,t)&=&\int d^d\vk \left(e^{-i\omega t}-\gamma^0 e^{i\omega t}\right)
 \tp^{(2)}(\vk)
 e^{i\vk.\vx}.
 \label{b2}
 \eea
 If one decomposes $\tp^{(a)}(\vk)$ as
 $\tp^{(a)}=\tp^{(a)}_++\tp^{(a)}_-$, $a=1,2$, where ,
 \be
 \tp^{(a)}_+=\frac{1+\gamma^0}{2}\tp^{(a)},\hspace{1cm}
 \tp^{(a)}_-=\frac{1-\gamma^0}{2}\tp^{(a)},
 \ee
 and using Eq.(\ref{b2}) one can show that the initial data $\p_0(\vx)=\p(\vx,0)$
 only determines $\tp^{(1)}_+$ and $\tp^{(2)}_-$:
 \be
 \tp^{(1)}_+(\vk)=\int d^d\vx e^{-i\vk.\vx}\p^+_0(\vx),\hspace{1cm}
 \tp^{(2)}_-(\vk)=\int d^d\vx e^{-i\vk.\vx}\p^-_0(\vx),
 \label{b3}
 \ee
 where $\p^\pm_0(\vx)=\frac{1\pm\gamma^0}{2}\p_0(\vx)$.  Defining,
 \be
 \chi^+(\vx)=\int d^d\vk e^{i\vk.\vx}\tp^{(1)}_-(\vk),\hspace{1cm}
 \chi^-(\vx)=\int d^d\vk e^{i\vk.\vx}\tp^{(2)}_+(\vk),
 \ee
 and using Eq.(\ref{b3}), one can determine $\p^{(a)}(\vx,t)$ in terms of
 fields $\p^\pm_0(\vx)$ and $\chi^\pm(\vx)$ living on the
 hypersurface $t=0$, as follows:
 \be
 \p^{(a)}(\vx,t)=\int d^d\vk d^d\vy\left(e^{-i\omega t}\pm\gamma^0 e^{i\omega
 t}\right)\left(\p^\pm_0(\vy)+\chi^\pm(\vy)\right)e^{i\vk.(\vx-\vy)}
 \label{b4}
 \ee
 Since $\gamma^0\p_0^\pm(\vx)=\pm\p_0^\pm(\vx)$, and
 $\gamma^0\chi^\pm=\mp\chi^\pm$,  one can decompose
 $\p(\vx,t)=\p^{(1)}(\vx,t)+\p^{(2)}(\vx,t)$ as
 eigenvectors of $\gamma^0$, say $\p_\pm$
 ($\gamma^0\p_\pm=\pm\p_\pm$), as follows,
 \bea
 \p(\vx,t)&=&\p_+(\vx,t)+\p_-(\vx,t)\nn\\
 \p_+(\vx,t)&=&\int d^d\vk d^d\vy\left[\left(e^{-i\omega t}+\gamma^0 e^{i\omega
 t}\right)\p^+_0(\vy)+\left(e^{-i\omega t}-\gamma^0 e^{i\omega
 t}\right)\chi^-(\vy)\right]e^{i\vk.(\vx-\vy)}\nn\\
 \p_-(\vx,t)&=&\int d^d\vk d^d\vy\left[\left(e^{-i\omega t}-\gamma^0
e^{i\omega t}\right)\p^-_0(\vy)+\left(e^{-i\omega t}+\gamma^0
e^{i\omega t}\right)\chi^+(\vy)\right]e^{i\vk.(\vx-\vy)}
 \label{GS}
 \eea
 As is shown in the appendix, $\chi^\pm$ can be determined in terms
 of  $\psi^\pm_0(\vx)$.
 \par
 Dirac equation can be obtained from the  Dirac action
 for spinors,
 \be
 S=i\int_{\Sigma} \pb{\overrightarrow{\not{\!\partial}}}\p.
 \ee
 if one assume that Dirac fields vanish on the boundary. This assumption is also necessary
 for Lagrangian density $i\pb{\overrightarrow{\not{\!\partial}}}\p$ to be Hermitian. If
 one rewrite the action as,
 \be
 S_D=i\int_{\Sigma} \pb\frac{{\overleftarrow{\not{\!\partial}}}+
 {\overrightarrow{\not{\!\partial}}}}{2}\p.
 \ee
 then the Lagrangian density is Hermitian with no condition on fields on the
 boundary. In order to obtain Dirac equation, one can also add the
 following  boundary term to the action,
 \be
 S_b=\int_{\partial \Sigma} \pb(\vx)\gamma^\mu\psi(\vx)n_\mu,
 \ee
 in which $n_\mu$ is the unit vector perpendicular to boundary. This term
 cancels the contribution from the variation of fields on the boundary to $\delta
 S$.  Therefore $\delta S=0$ gives the Dirac equation with no
 preassumption about fields on the boundary.
 \par
 Now assume that boundary is the hypesurface $t=0$ and insert $\p(\vx,t)$  the
 solution of the Dirac equation  into the action $S_D+S_b$. $S_D$
 vanishes by construction and,
 \be
 S_b=\int d^d\vx \pb(\vx)\gamma^0\p(\vx).
 \label{b5}
 \ee
 Using the identity,
 \be
 \p(\vx)=\int d^d\vk\int d^d\vy e^{i\vk.(\vx-\vy)}\p(\vy),
 \ee
 after some calculations, $S_b$ (\ref{b5}) can be given as follows,
 \be
 S_b=\mbox{Const.}\int d^d\vy d^d\vz\pb(\vy)\gamma^0\p(\vz) F(\vy-\vz),
 \ee
 where $F(\vx)=\int d^d\vk e^{i\vk.\vx}$. $F(\vx)$ is simply equal to $\delta^d(\vx)$.
 A representation of the Dirac delta function appropriate for our purposes can be
 obtained
 by noting that $F(\vx)=F({\bf R}.\vx)$ for any rotation
 ${\bf R}\in SO(d)$ and $F(s \vx)=s^{-d}F(\vx)$ for any $s>0$. Consequently,
 $$
 \vx.\nabla F(\vx)=r\frac{\partial}{\partial r}F(r)=-d\ F(\vx),
 $$
 where $r=\left|\vx\right|$.
 Therefore $F(\vx)=\mbox{Const.}
 \left|\vx\right|^{-d}$. Inserting $F(\vy-\vz)$ into $S_b$, one
 finally obtains,
 \be
 S_b=\mbox{Const.}\int d^d\vy d^d\vz
 \frac{\pb(\vz)\gamma^0\p(\vy)}{\left|\vy-\vz\right|^d}\nn\\
  \label{CFT}
 \ee
 %-----------------------------------------------------------------------
 \section{Spinors on $dS_{d+1}$}
 In the description of $dS_{d+1}$ as the domain $t>0$ with metric $ds^2=t^{-2}(dt^2-d\vx^2)$,
 the Klein-Gordon equation for massive scalar fields is,
 \be
 \left(t^2(\partial_t^2-\partial_\vx^2)+(1-d)t\partial_t+m^2\right)\phi=0.
 \label{c1}
 \ee
 If one write the above equation in the following form,
 \be
 \left(t^2(\partial_t^2-\partial_\vx^2)+(1-d)t\partial
 t+m^2\right)\p=
 \left(t\gamma^\mu\partial_\nu+\alpha\gamma^0+\zeta\right)
 \left(t\gamma^\nu\partial_\nu+\beta\gamma^0+\xi\right)\p,
 \ee
 where $\p$ is a spinor field, then one verify that
 \be
 \beta=1+\alpha=\frac{1-d}{2},\hspace{1cm}\xi=\zeta=0,\hspace{1cm}m^2=\frac{d^2-1}{4}.
 \ee
 Therefor, for fields with mass $m^2=\frac{d^2-1}{4}$, one can
 rewrite the Klein-Gordon equation (\ref{c1}), as
 \be
 \left(t\gamma^\mu\partial_\nu+\frac{1-d}{2}\gamma^0\right)\p=0.
 \label{c2}
 \ee
 As we observed in Eqs.(\ref{MKG}) and (\ref{dSKG}), massive scalar fields on $dS_{d+1}$ with
 mass $m^2=\frac{d^2-1}{4}$ are related to massless fields on $M_{d+1}$.
 It is easy to verify that if $\p$ is a solution of Dirac equation for massless spinors in
 $M_{d+1}$ then $\Psi=t^{\frac{d-1}{2}}\p$ is a solutions of Eq.(\ref{c2}). If one rewrite
 Eq.(\ref{c2}) in terms of the Dirac operator in $dS_{d+1}$,
 $\d=t\not{\!\partial}-\frac{d}{2}\gamma^0$, one obtains,
 \be
 \left(\d+\frac{1}{2}\gamma^0\right)\Psi=0.
 \ee
 Using the general solution of massless
 spinors in $M_{d+1}$ Eq.(\ref{GS}) one verifies that $\Psi_\pm=t^{\frac{d-1}{2}}\psi_\pm$
 are massive spinors in $dS_{d+1}$ with mass
 $m_\pm=\mp\frac{1}{2}$ respectively. In general, one can show
 that the most general solution of Dirac equation for spinors with mass
 $m$, $(\d-m)\Psi=0$, can be given in terms of $\p_\pm$, given in (\ref{GS}), as follows:
 \be
 \Psi=t^{\frac{d}{2}-m}\p_-+t^{\frac{d}{2}+m}\p_+.
 \ee
 In words, all massive spinors in $dS_{d+1}$ space can be considered as
 images of massless spinors in $M_{d+1}$. The extension of this
 result to massive spinors in Euclidean $AdS_{d+1}$ and massless
 spinors in $d+1$ dimensional Euclidean space is straightforward.
 \par
 By construction, the hypersurface $t=0$ in $M_{d+1}$ is isomorphic to the
 boundary of $dS_{d+1}$. If one identifies the CFT fields on the
 boundary of $dS_{d+1}$ with initial data $\p_0$ and $\chi$ on the
 hypersurface of $M_{d+1}$, then one can consider the action $S_b$
 given in Eq.(\ref{CFT}) as the action of CFT on the boundary of
 $dS_{d+1}$.
 \par
 The map $\psi_+\to\Psi_+=t^{\frac{d}{2}+m}\p_+$ can be used to
 find the Minkowski dual of certain fermionic vertices on dS
 space. Consider a vertex $V({\bar\Psi}_+\Psi_+)=g({\bar\Psi}_+\Psi_+)^n$
 in dS space where $n$ is some integer to be determined. The Dirac equation is
 \be
 (\d-m)\Psi_+=g{\bar{\Psi}}^n\Psi^{n-1},
 \label{d1}
 \ee
 in which $g$ is the coupling constant in dS space.
 Rewriting the above equation in terms of massless spinor $\p_+$,
 one obtains,
 \be
 \not\!\partial\p_+=g(t)\pb_+^n\p_+^{n-1},
 \ee
 where,
 \be
 g(t)=gt^{-(\frac{d}{2}+m+1)}t^{(2n-1)(\frac{d}{2}+m)},
 \ee
 is the coupling in the Minkowski spacetime.
 To obtain the above relation we have used the identity,
 $(\d-m)\Psi_+=t^{\frac{d}{2}+m+1}\not\!\partial\p_+$.  The Requirement that $g(t)$ is
 constant imposes the following identity:
 \be
 n=1+\frac{1}{d+2m}, \hspace{1cm}n\in  \mbox{N}
 \ee
 Therefore, $m=-\frac{d-1}{2}$ and $n=2$. One can verify that the field theory with
 $n=2$  vertex is renormalizable for $d\le 1$.
\par
 To study the differences between scalars and spinors in the map $M_{d+1}\to dS_{d+1}$,
 it is useful to search for those scalar field vertices that could
 be mapped from Minkowski spacetime to dS space. As we explained
 in section 1,
 massless scalar fields in $M_{d+1}$ can be only mapped to massive
 scalar fields in $dS_{d+1}$ with mass $m^2=\frac{d^2-1}{4}.$ Using the
 identity,
 \be
 \left(t^2\partial_t^2+(1-d)t\partial_t-t^2\nabla^2+\frac{d^2-1}{4}\right)\Phi(\vx,t)=
 t^{\frac{d-1}{2}+2}\left(\partial_t^2-\nabla^2\right)\phi,
 \ee
 in which $\Phi=t^{\frac{d-1}{2}}\phi$, and considering a vertex
 $V(\phi)=\phi^{\tilde n}$,
 one can show that
 \be
 {\tilde n}=2+\frac{4}{d-1}.
 \ee
 Since ${\tilde n}$ is an integer, the above identity can be
 satisfied only for special values of $d$: $d\in\{2,3,5\}$. The
 corresponding value for $\tilde n$ are $6,4,3$. All such theories
 are renormalizable.

 \section*{Summary} Studying the map between massless fields in
 $d+1$ dimensional Minkowski spacetime ($M_{d+1}$) and massive fields on
 de Sitter space $dS_{d+1}$, we found that $\p_\pm$, the massless scalars in
 $M_{d+1}$ which are eigenvectors of $\gamma^0$ with eigenvalues $\pm
 1$ can be mapped to massive spinors in $dS_{d+1}$ with mass,
 $m=\mp\frac{1}{2}$ respectively. We found that the conformal weight of dual spinors on the
 boundary of $dS_{d+1}$ space is $\Delta=\frac{d}{2}$ and the
 CFT action is
 \be
 S_b=\mbox{Const.}\int_{\partial \Sigma}
 \frac{\pb(\vz)\gamma^0\p(\vy)}{\left|\vy-\vz\right|^d},
 \ee
 which is in agreement with the action of scalar fields obtained in dS/CFT correspondence.
 We observed that  massive spinors on $dS_{d+1}$ with arbitrary
 mass can be given in terms of $\p_\pm$ as follows,
 \be
 \Psi=t^{\frac{d}{2}-m}\p_-+t^{\frac{d}{2}+m}\p_+.
 \ee
 Using the above identity, we found that the map between free
 field theories can be extended to theories with vertex
 $({\bar\p}\p)^2$ for spinors with mass $m=-\frac{d-1}{2}$ for arbitrary
 $d\ge1$.  In the case of scalar fields similar considerations showed
 that only for $d=2,3,5$, the $\phi^6$, $\phi^4$ and $\phi^3$ theories can be
 mapped from $M_{d+1}$ to $dS_{d+1}$ respectively.

%------------------------------------------------------------------
  \section*{Appendix}
 In this appendix we obtain the field $\chi(\vx)$ in terms of
 $\psi_0(\vx)$.  We also give the action $S_b$ in terms of $\p_0$ and ${\dot{\p}}_0$. The
 notation we use here for the Fourier components of fields differs
 slightly with those used in the main body of the paper.
 \par
 From Eq.(\ref{b4}) one can show that,
 \bea
 i{\dot\p}^\pm(\vx,0)&=&\int d^d\vy d^d\vk\
 \omega\chi^\pm(\vy)e^{i\vk.(\vx-\vy)}\nn\\
 &=&\int d^d\vk \ \omega
 {\tilde\chi}^\pm(\vk)e^{i\vk.\vx}.
 \eea
 On the other hand,
 $i{\dot\p}=-i\gamma^0\gamma^i\partial_i\psi$. Thus, using the
 identity $\p_0^\pm(\vx)=\int d^d\vk e^{i\vk.\vx}\tp^\pm_0(\vk)$, one verifies that
 \be
 \int d^d\vk\ \omega{\tilde\chi}^\pm(\vk)e^{i\vk.\vx}=
 \int d^d\vy d^d\vk\
 \gamma^0\gamma^ik_ie^{i\vk.\vx}\tp^\pm_0(\vk),
 \ee
 which results in the following identity:
 \bea
 {\tilde\chi}^\pm(\vk)&=&\gamma^0\gamma^i\frac{k_i}{\omega}\tp^\pm(\vk)\nn\\
 &=&\mp\gamma^i\frac{k_i}{\omega}\tp^\pm(\vk)
 \eea
 Using the above identity one can also show that
 \be
 \tp^\pm(\vk)=\mp\gamma^i\frac{k_i}{\omega}{\tilde\chi}^\pm(\vk)
 \label{ap1}
 \ee
 To action $S_b$ can be given in terms of $\p_0$ and ${\dot\p}_0$ as follows
 (see Eq.(\ref{b5})):
 \bea
 S_b&=&\int d^d\vx \pb(\vx)\gamma^0\p(\vx)\nn\\
 &=&\int d^d\vx
 \left(\pb^+(\vx)\p^+(\vx)-\pb^-(\vx)\p^-(\vx)\right)\nn\\
 &=&\int d^d\vk \left({\tilde{\pb}}^+(\vk)\tp^+(\vk)-
 {\tilde{\pb}}^-(\vk)\p^-(-\vk)\right)\nn\\
 &=&\int
 d^d\vk\ {\tilde{\pb}}^+(\vk)\tp^+(-\vk)+\int d^d\vk\
 {\tilde{\bar\chi}}^-(\vk){\tilde{\chi}}^-(-\vk)
 \eea
 where to obtain  the last equality we have used Eq.(\ref{ap1}).
 Using the Fourier transformation one obtains,
 \bea
 \int d^d\vk\ {\tilde{\pb}}^+(\vk)\tp^+(-\vk)&=&\int d^d\vy d^d\vz
 \pb^+(\vy)\p^+(\vz)\int d^d \vk\ e^{i\vk.(\vy-\vz)}\nn\\
 &=&\mbox{Const.}\int d^d\vy d^d\vz
 \frac{\pb^+(\vy)\p^+(\vz)}{\left|\vy-\vz\right|^d}.
 \eea
 The last equality can be obtained by considering the variation of
 the integrand under rescaling and rotation.
 \par
 Furthermore,
 \bea
 \int d^d\vk\ {\tilde{\bar\chi}}^-(\vk){\tilde{\chi}}^-(-\vk)&=&
 \int d^d\vk\
 \frac{1}{\omega^2}{\tilde{\dot{\pb}}}^-(\vk){\tilde{\dot{\pb}}}^-(-\vk)\nn\\
 &=&\int d^d\vy d^d\vz
 {\dot{\pb}}^-(\vy){\dot\p}^-(\vz)\int d^d \vk\
 \frac{1}{\omega^2}e^{i\vk.(\vy-\vz)}\nn\\
 &=&\mbox{Const.}\int d^d\vy d^d\vz
 \frac{{\dot{\pb}}^-(\vy){\dot\p}^-(\vz)}{\left|\vy-\vz\right|^{d-2}}.
 \eea
 This result is in agreement with those obtained in
 refs.\cite{map,Strom} for scalar fields.

\end{document}